# Spontaneous Bremsstrahlung in Scattering of an Electron by a Nucleus in the Field of Two Light Waves

S. P. Roshchupkin and O. B. Lysenko

*Institute of Applied Physics, National Academy of Sciences of Ukraine, Sumy, 40024 Ukraine*

We theoretically investigate nonresonant spontaneous bremsstrahlung in the scattering of an electron by a nucleus in the field of two linearly polarized light waves propagating in the same direction in the general relativistic case. It is demonstrated that there are two substantially different kinematic ranges: the noninterference range, where the Bunkin–Fedorov quantum parameters $\gamma_{1,2}$ (22) serve as multiphoton parameters, and the interference range, where interference effects become significant, and quantum interference parameters $\alpha_{\pm}$ (24) play the role of multiphoton parameters. We determine the cross sections of electron–nucleus spontaneous bremsstrahlung in these kinematic ranges. It is demonstrated that the partial cross section in the interference range with emission (absorption) of photons at combination frequencies may considerably exceed the corresponding cross section for any other geometry.

## 1. INTRODUCTION

Spontaneous bremsstrahlung (SB) accompanying electron–nucleus scattering in the field of a plane electromagnetic wave has been investigated for different ranges of the electron energy and the wave intensity for quite a long time (e.g., see [1–6]). These studies were mainly focused on the analysis of resonances that arise when the Green function of an intermediate electron reaches the mass shell in the field of a plane wave. In the general relativistic case, these resonances were considered in [7] (see also the review provided in [8]).

Recently, much attention has been given to the investigation of elementary quantum processes in the field of several laser waves. Specifically, Karapetyan and Fedorov [9] have studied stimulated bremsstrahlung emission and absorption (SBEA) accompanying electron–nucleus scattering in the presence of two plane electromagnetic waves in the nonrelativistic limit (within the framework of the dipole approximation). In the general relativistic case, this process was considered in [10–12] for the field of two waves and in [13] for the field of an arbitrary number of plane electromagnetic waves. These studies have revealed a physically new interference effect [11, 12] associated with emission and absorption of equal numbers of photons from both waves in electron–nucleus scattering by given angles in a certain plane (the interference effect was generalized to an arbitrary number of waves in [13]). The SBEA effect in electron–electron scattering in the field of two waves was studied in [14], while spontaneous emission of a photon by an electron in the field of two waves was considered earlier, e.g., in [15–19]. Within the interference range, this process was analyzed in [20] for the field of two waves and in [21] for the field of an arbitrary number of plane waves. These studies revealed a spontaneous combination effect implying that an electron emits a spontaneous photon at combination frequencies of applied waves.

In this paper, we consider nonresonant SB in electron–nucleus scattering in the field of two linearly polarized electromagnetic waves propagating in the same direction. We will thoroughly investigate electron–nucleus SB within the noninterference range and analyze the interference case, when electron scattering and spontaneous photon emission occur in the same plane perpendicular to the polarization vectors of the waves. Processes with emission of photons at combination frequencies become important in this regime. A relativistic system of units with $\hbar = c = 1$ will be employed in our analysis.

## 2. THE SB AMPLITUDE

Let us choose the four-potential of the external field in the form of a sum of two linearly polarized electromagnetic waves propagating along the $z$-axis:

$$A = A_1(\varphi_1) + A_2(\varphi_2), \qquad (1)$$

where

$$A_j(\varphi_j) = \frac{F_j}{\omega_j}(e_j \cos\varphi_j). \qquad (2)$$

Here, $e_j = (0, \mathbf{e}_j)$ is the four-vector of wave polarization ($\mathbf{e}_1$ and $\mathbf{e}_2$ lie in the $xy$ plane, and $\Delta = \angle(\mathbf{e}_1, \mathbf{e}_2)$); $F_j$ and $\omega_j$ are the strength and the frequency of the first ($j = 1$) and the second ($j = 2$) wave, respectively; and the argument $\varphi_j$ is written as

$$\varphi_j = \omega_j(t - z), \quad j = 1, 2. \quad (3)$$

The electron–nucleus SB amplitude in the light field (1) is given by

$$S_{fi} = -ie^2 \int d^4x_1 d^4x_2 \overline{\psi}_f(x_2|A)[\tilde{\gamma}_0 A_0(x_2) G(x_1 x_2|A) \\ \times \hat{A}'(x_1, k') + \hat{A}'(x_2, k') G(x_1 x_2|A) \tilde{\gamma}_0 A_0(x_1)] \psi_i(x_1|A), \quad (4)$$

where $\psi_i(x_1|A)$ and $\overline{\psi}_f(x_2|A)$ are the wave functions (Volkov functions) of an electron in the initial and final states in the wave field (1) [22]; $G(x_1, x_2|A)$ is the Green function of an electron in the field of a plane wave (1) [7, 23–26];

$$A_0(x) = \frac{Ze}{|\mathbf{x}|} \quad (5)$$

is the Coulomb potential of the nucleus $Ze$; and $\hat{A}'(x, k') = \tilde{\gamma}^\mu A'_\mu$ denotes the scalar product of Dirac matrices $\tilde{\gamma}^\mu$ ($\mu = 0, 1, 2, 3$) and the four-potential of the spontaneous photon,

$$A'_\mu(x, k') = \sqrt{\frac{2\pi}{\omega'}} \varepsilon^*_\mu \exp(ik'x). \quad (6)$$

Here, $\varepsilon^*_\mu$ is the polarization four-vector and $k' = \omega' n' = \omega'(1, \mathbf{n}')$ is the four-momentum of the spontaneous photon. Performing simple calculations (see [7, 22]), we can reduce expression (4) for the SB amplitude to the following form:

$$S_{fi} = \sum_{l=-\infty}^{\infty} \sum_{s=-\infty}^{\infty} S_{ls}, \quad (7)$$

where the partial amplitude corresponding to emission ($l > 0$, $s > 0$) and absorption ($l < 0$, $s < 0$) of $|l|$ photons from the first wave and $|s|$ photons from the second wave is given by

$$S_{ls} = -i\frac{8\pi^{5/2}Ze^3}{\sqrt{2\omega' \tilde{E}_i \tilde{E}_f}} \exp(i\phi_{if}) [\bar{u}_f H_{ls} u_i] \frac{\delta(q_0)}{\mathbf{q}^2}, \quad (8)$$

$$H_{ls} = \sum_{l'=-\infty}^{\infty} \sum_{s'=-\infty}^{\infty} \left[ M_{l-l', s-s'}(\tilde{p}_f, \tilde{q}_i) \frac{\hat{\tilde{q}}_i + m_*}{\tilde{q}_i^2 - m_*^2} K_{l', s'}(\tilde{q}_i, \tilde{p}_i) \right. \\ \left. + K_{l', s'}(\tilde{p}_f, \tilde{q}_f) \frac{\hat{\tilde{q}}_f + m_*}{\tilde{q}_f^2 - m_*^2} M_{l-l', s-s'}(\tilde{q}_f, \tilde{p}_i) \right]. \quad (9)$$

Here, $\phi_{i,f}$ is the phase that is independent of summation indices; $u_i$ and $\bar{u}_f$ are the Dirac bispinors; $q = (q_0, \mathbf{q})$ is the transferred four-momentum; and $\tilde{q}_i$ and $\tilde{q}_f$ are the four-momenta of intermediate electrons for the forward and exchange amplitudes,

$$q = \tilde{p}_f - \tilde{p}_i + k' + lk_1 + sk_2, \quad (10)$$

$$\tilde{q}_i = \tilde{p}_i - k' - l'k_1 - s'k_2, \quad (11)$$

$$\tilde{q}_f = \tilde{p}_f + k' + l'k_1 + s'k_2, \quad (12)$$

where $k_1 = \omega_1 n = \omega_1(1, \mathbf{n})$ and $k_2 = \omega_2 n = \omega_2(1, \mathbf{n})$ are the four-momenta of photons from the first and second waves, respectively; $\tilde{p}_i$ and $\tilde{p}_f$ are the four-quasimomenta of the electron before and after scattering; and $m_*$ is the effective mass of the electron in the wave field (1),

$$\tilde{p}_j = p_j + \frac{m^2}{4(k_1 p_j)}(\eta_1^2 + \eta_2^2) k_1, \quad j = i, f. \quad (13)$$

$$m_* = m\sqrt{1 + \frac{1}{2}\eta_1^2 + \frac{1}{2}\eta_2^2}. \quad (14)$$

Here, $p_j = (E_j, \mathbf{p}_j)$ is the four-momentum of the electron before ($j = i$) and after ($j = f$) scattering and

$$\eta_{1,2} = \frac{eF_{1,2}}{m\omega_{1,2}} \quad (15)$$

is the classical relativistic-invariant parameter that characterizes the intensities of the first and second waves.

Operators $M_{rr'}$ in expression (9), which determine the amplitude of electron–nucleus scattering (if the intermediate electron is real, see [11]) in the field of two waves, are written as

$$M_{rr'}(\tilde{p}_2, \tilde{p}_1) = \tilde{\gamma}_0 I_{rr'} + \frac{\omega_1 m^2}{8(k_1 \tilde{p}_1)(k_1 \tilde{p}_2)} B_{rr'} \hat{k}_1 \\ + \frac{m}{4(k_1 \tilde{p}_1)} \tilde{\gamma}_0 \hat{k}_1 \hat{D}_{rr'} + \frac{m}{4(k_1 \tilde{p}_2)} \hat{D}_{rr'} \hat{k}_1 \tilde{\gamma}_0, \quad (16)$$

where $r = l - l'$, $r' = s - s'$, and $\tilde{p}_1 = \tilde{q}_i$ and $\tilde{p}_2 = \tilde{p}_f$ for the forward amplitude and $\tilde{p}_1 = \tilde{p}_i$ and $\tilde{p}_2 = \tilde{q}_f$ for the exchange amplitude. Operators $K_{l's'}$ in (9), which determine the amplitude of spontaneous emission of a photon by an electron (if the intermediate electron is real, see [20]) in the field of two waves, are given by

$$K_{l's'}(\tilde{p}_2, \tilde{p}_1) = \hat{\varepsilon}^* I_{l's'} + \frac{\omega_1 m^2}{8(k_1 \tilde{p}_1)(k_1 \tilde{p}_2)} B_{l's'} \hat{k}_1 \\ + \frac{m}{4(k_1 \tilde{p}_1)} \hat{\varepsilon}^* \hat{k}_1 \hat{D}_{l's'} + \frac{m}{4(k_1 \tilde{p}_2)} \hat{D}_{l's'} \hat{k}_1 \hat{\varepsilon}^*, \quad (17)$$

where $\tilde{p}_1 = \tilde{p}_i$ and $\tilde{p}_2 = \tilde{q}_i$ for the forward amplitude and $\tilde{p}_1 = \tilde{q}_f$ and $\tilde{p}_2 = \tilde{p}_f$ for the exchange amplitude. Functions $I_{rr'}$ and $B_{rr'}$ and the four-vector $D_{rr'}$ in expres-

sions (16) and (17) were first introduced in [11] (see also [12, 13]) and are written as

$$D_{rr'} = e_1\eta_1(I_{r+1,r'} + I_{r-1,r'}) \\ + e_2\eta_2(I_{r,r'+1} + I_{r,r'-1}), \quad (18)$$

$$B_{rr'} = \eta_1^2[I_{r+2,r'} + I_{r-2,r'} + 2I_{rr'}] \\ + \eta_2^2[I_{r,r'+2} + I_{r,r'+2} + 2I_{rr'}] + 2\eta_1\eta_2 \quad (19) \\ \times [I_{r-1,r'-1} + I_{r+1,r'+1} + I_{r-1,r'+1} + I_{r+1,r'-1}]\cos\Delta.$$

The functions $I_{rr'}$ can be represented as series expansions in integer-order Bessel functions $J_r$ [11]:

$$I_{rr'}(\gamma_1, \beta_1; \gamma_2, \beta_2; \alpha_+, \alpha_-) \\ = \sum_{j=-\infty}^{\infty}\sum_{j'=-\infty}^{\infty} J_j(\alpha_+)J_{j'}(\alpha_-)J_{r-j-j'}(\gamma_1, \beta_1)J_{r'+j'-j}(\gamma_2, \beta_2). \quad (20)$$

The generalized Bessel functions $J_r(\gamma, \beta)$, which describe multiphoton processes in the field of a single wave [11, 27–29], were thoroughly studied by Reiss [30]:

$$J_r(\gamma, \beta) = \sum_{r'=-\infty}^{\infty} J_{r-2r'}(\gamma)J_{r'}(\beta). \quad (21)$$

The arguments of the functions $I_{rr'}$ are given by

$$\gamma_j \equiv \gamma_j(\tilde{p}_2, \tilde{p}_1) = m\eta_j|e_jg_j|, \\ g_j \equiv g_j(\tilde{p}_2, \tilde{p}_1) = \frac{\tilde{p}_2}{(k_j\tilde{p}_2)} - \frac{\tilde{p}_1}{(k_j\tilde{p}_1)}, \quad (22)$$

$$\beta_j \equiv \beta_j(\tilde{p}_2, \tilde{p}_1) = \frac{1}{8}\eta_j^2 m^2\left[\frac{1}{(k_j\tilde{p}_2)} - \frac{1}{(k_j\tilde{p}_1)}\right], \quad (23) \\ j = 1, 2;$$

$$\alpha_\pm \equiv \alpha_\pm(\tilde{p}_2, \tilde{p}_1) = \eta_1\eta_2\frac{m^2|\cos\Delta|}{2(k_1 \pm k_2)}\left(\frac{1}{\tilde{p}_1} - \frac{1}{\tilde{p}_2}\right). \quad (24)$$

We emphasize that the values of integer indices $r$ and $r'$ in formulas (18)–(20) and the values of four-momenta $\tilde{p}_1$ and $\tilde{p}_2$ in (18), (19), and (21)–(24) are equal to the relevant values of the quantities involved in (16) and (17) (see the text below these formulas). Note that the quantity $\gamma_j$ (22) is the well-known Bunkin–Fedorov quantum multiphoton parameter [27–29]. Quantum parameters $\beta_j$ (23) play an important role for high electron energies ($\beta_j = 0$ in the dipole approximation in the interaction of an electron with electric fields in both waves). Parameters $\alpha_\pm$ (24) are the quantum interference parameters, which govern interference effects in electron–nucleus scattering and spontaneous emission of a photon by an electron in the field of two waves. We should note also that the functions $I_{rr'}$ (20) involved in (16) and (17) govern multiphoton processes in the field of two waves. These functions were thoroughly studied in [11]. Here, we should mention only that, if the quantum interference parameters satisfy the relation $\alpha_\pm \gtrsim 1$, then the processes of correlated emission and absorption of equal numbers of photons from both waves become significant. In other words, the interaction of an electron with the fields of both waves is effectively reduced to the emission or absorption of photons at combination frequencies $|\omega_1 \pm \omega_2|$. If the quantum interference parameters satisfy the inequality $\alpha_\pm \ll 1$ (this condition is met in the dipole approximation; for moderate wave intensities, $\eta_1\eta_2 \ll \omega_{1,2}E_i/(m^2v_i)$, with arbitrary electron energies within the optical frequency range; and for arbitrary wave intensities in cases where the angle between the polarization vectors is $\Delta = \pi/2$, see (24) and [11, 12]), then we can neglect the influence of interference processes ($j = j' = 0$). In this case, the functions $I_{rr'}$ (20) can be written as products of generalized Bessel functions corresponding to independent emission and absorption of photons from the first and second waves:

$$I_{rr'}(\gamma_1, \beta_1; \gamma_2, \beta_2; 0, 0) = J_r(\gamma_1, \beta_1)J_{r'}(\gamma_2, \beta_2). \quad (25)$$

Expressions (7)–(9) for the electron–nucleus SB amplitude are valid for arbitrary intensities and frequencies of both waves and for electron velocities $v_{i,f} \gg Z/137$. It can be easily shown that, when one of the waves is switched off (e.g., with $F_2 = 0$), formulas (7)–(9) define the electron–nucleus SB amplitude in the field of a single wave [7]. When both waves are switched off ($F_1 = F_2 = 0$), these formulas describe the conventional electron–nucleus SB amplitude in the absence of an external field [22].

When the frequencies of the light waves are equal to each other ($\omega_1 = \omega_2$), expressions (7)–(9) should be reduced to a formula for the electron–nucleus SB amplitude in the field of a single wave whose strength $F$ and polarization vector $\mathbf{e}$ are related to the initial parameters of the light waves by the following expressions

$$F = \sqrt{F_1^2 + F_2^2 + 2F_1F_2\cos\Delta}, \\ \mathbf{e} = \frac{1}{F}(F_1\mathbf{e}_1 + F_2\mathbf{e}_2). \quad (26)$$

Introducing the difference between the frequencies of the light waves, $\Delta\omega = \omega_2 - \omega_1$, we can rewrite expressions (10)–(12) for the four-momenta as

$$q = \tilde{p}_f - \tilde{p}_i + k' + [l + s(1 - \Delta\omega/\omega_1)]k_1, \quad (27)$$

$$\tilde{q}_i = \tilde{p}_i - k' - [l' + s'(1 + \Delta\omega/\omega_1)]k_1, \quad (28)$$

$$\tilde{q}_f = \tilde{p}_f + k' + [l' + s'(1 + \Delta\omega/\omega_1)]k_1, \quad (29)$$

As can be seen from (27)–(29), if

$$|\Delta\omega|/\omega_1 \ll 1, \quad (30)$$

this term can be neglected. Then, we can introduce new photon numbers, $r = l + s$ and $r' = l' + s'$, and perform summation in expressions (7)–(9) for the SB amplitude first in all values of $s$ and then in all values of $s'$. One can easily verify that

$$\sum_{s=-\infty}^{\infty} I_{r-r'+s'-s,\, s-s'} = J_{r-r'}(\gamma, \beta),$$

$$\sum_{s'=-\infty}^{\infty} I_{r'-s',\, s'} = J_{r'}(\gamma', \beta'), \quad (31)$$

where

$$\gamma = \sqrt{\gamma_1^2 + \gamma_2^2 + 2\gamma_1\gamma_2 \cos\Delta}, \quad (32)$$

$$\beta = \beta_1 + \beta_2 + \alpha_+$$
$$= \frac{1}{8} m^2 \left[ \frac{1}{(k_1 \tilde{p}_2)} - \frac{1}{(k_1 \tilde{p}_1)} \right] (\eta_1^2 + \eta_2^2 + 2\eta_1\eta_2 \cos\Delta). \quad (33)$$

Note that parameters with primes in (31) differ from the relevant parameters without primes by the values of the four-momenta $\tilde{p}_1$ and $\tilde{p}_2$ [see the text below formulas (16) and (17)]. Hence, we derive

$$\sum_{s=-\infty}^{\infty} M_{r-r'+s'-s,\, s-s'} \longrightarrow M_{r-r'},$$

$$\sum_{s'=-\infty}^{\infty} K_{r'-s',\, s'} \longrightarrow K_{r'}, \quad (34)$$

which brings us to the final expression

$$\sum_{s=-\infty}^{\infty} H_{r-s,\, s} \longrightarrow H_r, \quad (35)$$

where the quantities $M_{r-r'}$ and $K_{r'}$ determine the amplitudes of electron–nucleus scattering and spontaneous emission of a photon by an electron in the field of a single wave. With allowance for (8), the quantity $H_r$ determines the electron–nucleus SB amplitude in the field of a single wave [7, 8]. Thus, with inequality (30), we arrive at a thoroughly studied electron–nucleus SB process in the field of a plane wave [7]. Therefore, in what follows, we assume that the frequencies of the light waves are not close to each other, i.e., consider a situation opposite to (30)

$$|\Delta\omega|/\omega_1 \gtrsim 1. \quad (36)$$

In addition, we assume that the frequencies of the light waves satisfy the following conditions

$$\omega_1 > \omega_2, \quad \omega_{1,2} \ll \begin{cases} m, & \text{if } E_i \gtrsim m \\ m v_i^2/2, & \text{if } v_i \ll 1. \end{cases} \quad (37)$$

As can be seen from expression (22), the Bunkin–Fedorov parameters $\gamma_j$ are highly sensitive to the kinematics of electron scattering and spontaneous photon emission. Let us assume that the polarization vectors of both waves are directed along the $x$-axis ($\Delta = 0$):

$$\mathbf{e}_1 = \mathbf{e}_2 = \mathbf{e}_x. \quad (38)$$

In this case, if electron scattering and spontaneous photon emission occur in the same plane, perpendicular to the polarization vector ($\mathbf{e}_x \mathbf{k'} = \mathbf{e}_x \mathbf{p}_i = \mathbf{e}_x \mathbf{p}_f = 0$), then the quantum parameters are equal to zero, $\gamma_j = 0$, and multiphoton processes are completely determined by quantum parameters $\beta_j$ (23) and $\alpha_\pm$ (24). This range of parameters will be referred to as the interference range. Within the interference range, the Bunkin–Fedorov quantum parameters are small for arbitrary intensities of both waves if the angle between the electron scattering plane and the polarization vector ($\varphi = \angle[(\mathbf{p}_i, \mathbf{p}_f), \mathbf{e}_x]$) and the angle between the plane defined by the momentum of the spontaneous photon and the initial electron and the polarization vector ($\psi = \angle[(\mathbf{p}_i, \mathbf{k'}), \mathbf{e}_x]$) are close to $\pi/2$:

$$\left|\varphi - \frac{\pi}{2}\right| \ll \frac{\omega_{1,2}}{m v_i \eta_{1,2}} \leq 1,$$

$$\left|\psi - \frac{\pi}{2}\right| \ll \frac{\omega_{1,2}}{m v_i \eta_{1,2}} \leq 1. \quad (39)$$

The noninterference range is defined by the inequalities opposite to (39):

$$\left|\varphi - \frac{\pi}{2}\right| \gtrsim \frac{\omega_{1,2}}{m v_i \eta_{1,2}} \leq 1,$$

$$\left|\psi - \frac{\pi}{2}\right| \gtrsim \frac{\omega_{1,2}}{m v_i \eta_{1,2}} \leq 1. \quad (40)$$

In this range, Bunkin–Fedorov quantum parameters play the role of the main multiphoton parameters. We emphasize that, within the interference range (39), we have $\gamma_j = 0$, and functions $I_{rr'}$ (20), which determine the SB amplitude through (7)–(9), are reduced to the functions $J_{r_1 r_2}$ [11]:

$$J_{r_1 r_2}(\beta_1, \beta_2; \alpha_+, \alpha_-) \equiv I_{rr'}(0, \beta_1; 0, \beta_2; \alpha_+, \alpha_-)$$
$$= \sum_{j=-\infty}^{\infty} \sum_{j'=-\infty}^{\infty} J_{r_1-j-j'}(\alpha_+) J_{r_2-j+j'}(\alpha_-) J_j(\beta_1) J_{j'}(\beta_2), \quad (41)$$

where

$$r_1 = \frac{1}{2}(r + r'), \quad r_2 = \frac{1}{2}(r - r'). \quad (42)$$

Hence, within the interference range, the numbers of photons emitted and absorbed by an electron from both waves correlate with each other in such a way that the half-sum and the half-difference of these numbers ($r$ and $r'$) are integers ($r_1$ and $r_2$). In what follows, we will sequentially consider electron–nucleus SB within the ranges defined by (40) and (39).

## 3. ELECTRON–NUCLEUS SB WITHIN THE NONINTERFERENCE RANGE

Let us consider electron–nucleus SB within the noninterference range defined by (40), i.e., within a kinematic range where the Bunkin–Fedorov quantum parameters $\gamma_j$ (22) are not small and are the main multiphoton parameters. Note that this range is rather broad. Within the noninterference range, quantum parameters (22)–(24) have the following orders of magnitude:

$$\gamma_{1,2} \sim \eta_{1,2} \frac{m v_{i,f}}{\omega_{1,2}}, \quad \beta_{1,2} \sim \gamma_{1,2} \xi_{1,2}, \quad (43)$$
$$\alpha_\pm \sim \gamma_1 \xi_2 \sim \gamma_2 \xi_1,$$

where

$$\xi_{1,2} = \eta_{1,2} \frac{m}{|\mathbf{p}_{i,f}|} \quad (44)$$

is the classical parameter that determines the integral characteristics of the process under study within the noninterference range [8, 11–13]. For moderately strong fields, we have $\xi_{1,2} \ll 1$, which is equivalent to the following restrictions on field intensities as functions of the electron energy:

$$\eta_{1,2} \ll \begin{cases} v_{i,f}, & \text{if } v_{i,f} \ll 1 \\ 1, & \text{if } E_{i,f} \sim m \\ E_{i,f}/m, & \text{if } E_{i,f} \gg m. \end{cases} \quad (45)$$

When conditions (45) are satisfied, we have $\beta_{1,2} \ll \gamma_{1,2}$ and $\alpha_\pm \ll \gamma_{1,2}$, and multiphoton processes are mainly determined by the Bunkin–Fedorov quantum parameters ($l \lesssim \gamma_1$ and $s \lesssim \gamma_2$). Taking into account that $l\omega_1/E_i \lesssim \xi_1 \ll 1$ and $l\omega_2/E_i \lesssim \xi_2 \ll 1$, we find that, within the range of moderately strong fields (45), expressions (8)–(17) for the SB amplitude can be considerably simplified. In particular, expressions (10)–(12) for the four-momenta and formulas (16) and (17) for the amplitudes can be rewritten as

$$q = p_f - p_i + k', \quad q_i = p_i - k', \quad q_f = p_f + k', \quad (46)$$

$$M_{l-l', s-s'} = \tilde{\gamma}_0 I_{l-l', s-s'}, \quad K_{l', s'} = \hat{\varepsilon}^* I_{l', s'}. \quad (47)$$

As can be seen from (46), within the range of moderately strong fields, no resonances occur when the Green function of an intermediate electron in the field of light waves reaches the mass shell ($q_i^2 \neq 0$ and $q_f^2 \neq 0$). In other words, the range of fields specified by (45) also defines the nonresonant range.

Since the arguments of the functions $I_{l', s'}$ and $I_{l-l', s-s'}$ involved in (47) are independent of summation indices, we can easily perform summation over all the integer indices $l'$ and $s'$ in the amplitude (9) with allowance for (47). The final expression for the SB amplitude defined by (8) and (9) is written as

$$S_{ls} = I_{ls} S_*, \quad (48)$$

where $S_*$ is the electron–nucleus SB amplitude in the absence of an external field [22]. The functions $I_{ls}$ are defined by expressions (20) and (21) where the arguments (22)–(24) are taken with $\tilde{p}_1 = p_i$ and $\tilde{p}_2 = p_f$, i.e.,

$$\gamma_{1,2} = \gamma_{1,2}(p_f, p_i), \quad \beta_{1,2} = \beta_{1,2}(p_f, p_i), \quad (49)$$
$$\alpha_\pm = \alpha_\pm(p_f, p_i).$$

Taking into account formula (48) for the amplitude, we can immediately derive an expression for the partial differential cross section of electron–nucleus SB in the field of two waves:

$$d\sigma_{ls} = |I_{ls}|^2 d\sigma_*. \quad (50)$$

Here, $d\sigma_*$ is the differential cross section of electron–nucleus SB in the absence of an external field [22].

As can be seen from (50), within the range of moderately strong fields (45), the cross section of electron–nucleus SB can be factorized as a product of the probability to emit (absorb) $l$ photons of the first wave and $s$ photons of the second wave and the electron–nucleus SB cross section in the absence of an external field. In other words, a spontaneous photon is emitted in this case independently of the emission (absorption) of photons from both waves through electron–nucleus bremsstrahlung. Generally, emission and absorption of photons from the first and second waves correlate with each other in such a situation through quantum interference parameters $\alpha_\pm$.

Consider the case when the Bunkin–Fedorov quantum parameters satisfy inequalities $\gamma_{1,2} \ll \xi_{1,2}^{-1} \gg 1$ and $\gamma_{1,2} \ll \xi_{2,1}^{-1} \gg 1$, which are equivalent to the following restrictions on the field intensities as functions of the electron energy:

$$\eta_{1,2}^2 \ll \begin{cases} \omega_{1,2}/m v_{i,f}, & \text{if } v_{i,f} \ll 1 \\ \omega_{1,2}/m, & \text{if } E_{i,f} \sim m \\ \omega_{1,2} E_{i,f}/m^2, & \text{if } E_{i,f} \gg m \end{cases},$$

$$\eta_1 \eta_2 \ll \begin{cases} \omega_{1,2}/m v_{i,f}, & \text{if } v_{i,f} \ll 1 \\ \omega_{1,2}/m, & \text{if } E_{i,f} \sim m \\ \omega_{1,2} E_{i,f}/m^2, & \text{if } E_{i,f} \gg m. \end{cases} \quad (51)$$

In such a situation, we have $\beta_{1,2} \ll 1$ and $\alpha_\pm \ll 1$ [see (43)]. Therefore, the functions $I_{ls}$ involved in (50) and defined by (20) and (21) are split into a product of inde-pendent Bessel functions [see (29)]. Then, the partial cross section (55) is written as

$$d\sigma_{ls} = J_l^2(\gamma_1)J_s^2(\gamma_2)d\sigma_*. \qquad (52)$$

Thus for intensities satisfying conditions (51), photons of the first and second waves are emitted (absorbed) independently of the emission (absorption) of a spontaneous photon through electron–nucleus bremsstrahlung. We emphasize that conditions (51) are much more stringent than conditions (45) (within the range of optical frequencies, $\omega_{1,2}/m \sim 10^{-5}$) and are equivalent to the dipole approximation in the limiting case of nonrelativistic energies.

The partial cross sections (50) and (52) can be readily summed over all possible processes of absorption and emission of photons from both waves. A natural result of this procedure is that the total cross section coincides with the electron–nucleus SB cross section in the absence of an external field. In other words, all the essentially quantum terms cancel each other out after summation, similar to the case of a single wave [7],

$$d\sigma = \sum_{l=-\infty}^{\infty}\sum_{s=-\infty}^{\infty} d\sigma_{ls} = d\sigma_*. \qquad (53)$$

## 4. ELECTRON–NUCLEUS SB IN THE INTERFERENCE RANGE

Let us consider electron–nucleus SB within the interference range, i.e., in the range where conditions (38) and (39) are satisfied. All the four Bunkin–Fedorov quantum parameters are small within this range, and the functions $I_{rr'}$ (20) are reduced to the functions $J_{r_1r_2}$ (41). Let us rewrite the SB amplitude defined by (7)–(12) and (16)–(19) taking into account new notations of summation indices (42). Then, the sum and the difference of integer indices $l'$, $s'$ and $l$, $s$ can be either even,

$$\begin{cases} l' + s' = 2s_1 \\ l' - s' = 2s_2 \end{cases}, \quad \begin{cases} l + s = 2l_1 \\ l - s = 2l_2 \end{cases}, \qquad (54)$$

or odd,

$$\begin{cases} l' + s' + 1 = 2s_1 \\ l' - s' + 1 = 2s_2 \end{cases}, \quad \begin{cases} l + s + 1 = 2l_1 \\ l - s + 1 = 2l_2 \end{cases}. \qquad (55)$$

Formulas (54) correspond to the emission (absorption) of an integer number of photons at combination frequencies ($\omega_1 \pm \omega_2$), when the four-momenta (10)–(12) are given by

$$\begin{cases} q \longrightarrow q' = \tilde{p}_f - \tilde{p}_i - k' + l_1(k_1 + k_2) + l_2(k_1 - k_2) \\ \tilde{q}_i \longrightarrow \tilde{q}'_i = \tilde{p}_i - k' - s_1(k_1 + k_2) - s_2(k_1 - k_2) \\ \tilde{q}_f \qquad \tilde{q}'_f = \tilde{p}_f + k' + s_1(k_1 + k_2) + s_2(k_1 - k_2). \end{cases} \qquad (56)$$

In this case, we have $I_{l's'} \longrightarrow J_{s_1 s_2}$, $I_{l' \pm 2, s'} \longrightarrow J_{s_1 \pm 1, s_2 \pm 1}$, $I_{l', s' \pm 2} \longrightarrow J_{s_1 \pm 1, s_2 \mp 1}$, $I_{l' \pm 1, s' \pm 1} \longrightarrow J_{s_1 \pm 1, s_2}$, $I_{l' \pm 1, s' \mp 1} \longrightarrow J_{s_1 s_2 \pm 1}$, and $D_{l's'} \longrightarrow 0$ [the four-vector $D_{l's'}$ differs from zero only when conditions (55) are met]. In a situation described by formulas (55), expressions (56) should be replaced by

$$\begin{cases} q \longrightarrow q'' = \tilde{p}_f - \tilde{p}_i - k' + \left(l_1 - \frac{1}{2}\right)(k_1 + k_2) \\ \qquad + \left(l_2 - \frac{1}{2}\right)(k_1 - k_2) \\ \tilde{q}_i \longrightarrow \tilde{q}''_i = \tilde{p}_i - k' - \left(s_1 - \frac{1}{2}\right)(k_1 + k_2) \\ \qquad - \left(s_2 - \frac{1}{2}\right)(k_1 - k_2) \\ \tilde{q}_f \longrightarrow \tilde{q}''_f = \tilde{p}_f + k' + \left(s_1 - \frac{1}{2}\right)(k_1 + k_2) \\ \qquad + \left(s_2 - \frac{1}{2}\right)(k_1 - k_2). \end{cases} \qquad (57)$$

In this case, we have $I_{l'+1, s'} \longrightarrow J_{s_1 s_2}$, $I_{l', s'+1} \longrightarrow J_{s_1, s_2-1}$, $I_{l'-1, s'} \longrightarrow J_{s_1-1, s_2-1}$, $I_{l', s'-1} \longrightarrow J_{s_1-1, s_2}$, $I_{l's'} \longrightarrow 0$, and $B_{l's'} \longrightarrow 0$. As can be seen from (56) and (57), within the interference range, an electron emits (absorbs) an integer or a half-integer number of photons at combination frequencies ($\omega_1 \pm \omega_2$) in the process of electron–nucleus bremsstrahlung and emission of a spontaneous photon. We emphasize that emission (absorption) of a half-integer number of photons at combination frequencies, in fact, implies that emission or absorption of an integer number of photons at combination frequencies is accompanied by emission or absorption of yet another photon with a frequency $\omega_1$ or $\omega_2$. The final expression for the electron–nucleus SB amplitude in the interference range is written as

$$S_{fi} = \sum_{l_1=-\infty}^{\infty}\sum_{l_2=-\infty}^{\infty} [S_{l_1 l_2}^{\text{int}} + S_{l_1 l_2}^{\text{half-ind}}], \qquad (62)$$

where $S_{l_1 l_2}^{\text{int}}$ and $S_{l_1 l_2}^{\text{half-ind}}$ are the partial amplitudes corresponding to the emission (absorption) of an integer and a half-integer number of photons at combination frequencies, respectively. These amplitudes are determined by expression (8) where the replacements $q \to q'$, $H_{ls} \to H^{\text{int}}_{l_1 l_2}$ and $q \to q''$, $H^{\text{half-ind}}_{l_1 l_2}$ should be made. Thus, we have

$$H^{\text{int}}_{l_1 l_2} = \sum_{s_1 = -\infty}^{\infty} \sum_{s_2 = -\infty}^{\infty} \left[ M'_{l_1 - s_1, l_2 - s_2}(\tilde{p}_f, \tilde{q}'_i) \frac{\hat{\tilde{q}}'_i + m_*}{\tilde{q}'^2_i - m^2_*} \right.$$
$$\times K'_{s_1 s_2}(\tilde{q}'_i, \tilde{p}_i) + K'_{s_1 s_2}(\tilde{p}_f, \tilde{q}'_f) \frac{\hat{\tilde{q}}'_f + m_*}{\tilde{q}'^2_f - m^2_*}$$
$$\times M'_{l_1 - s_1, l_2 - s_2}(\tilde{q}'_f, \tilde{p}_i) + M''_{l_1 - s_1 + 1, l_2 - s_2 + 1}(\tilde{p}_f, \tilde{q}''_i) \quad (63)$$
$$\times \frac{\hat{\tilde{q}}''_i + m_*}{\tilde{q}''^2_i - m^2_*} K''_{s_1 s_2}(\tilde{q}''_i, \tilde{p}_i) + K''_{s_1 s_2}(\tilde{p}_f, \tilde{q}''_f)$$
$$\left. \times \frac{\hat{\tilde{q}}''_f + m_*}{\tilde{q}''^2_f - m^2_*} M''_{l_1 - s_1 + 1, l_2 - s_2 + 1}(\tilde{q}''_f, \tilde{p}_i) \right],$$

$$H^{\text{half-ind}}_{l_1 l_2} = \sum_{s_1 = -\infty}^{\infty} \sum_{s_2 = -\infty}^{\infty} \left[ M''_{l_1 - s_1, l_2 - s_2}(\tilde{p}_f, \tilde{q}'_i) \frac{\hat{\tilde{q}}'_i + m_*}{\tilde{q}'^2_i - m^2_*} \right.$$
$$\times K'_{s_1 s_2}(\tilde{q}'_i, \tilde{p}_i) + K'_{s_1 s_2}(\tilde{p}_f, \tilde{q}'_f) \frac{\hat{\tilde{q}}'_f + m_*}{\tilde{q}'^2_f - m^2_*}$$
$$\times M''_{l_1 - s_1, l_2 - s_2}(\tilde{q}'_f, \tilde{p}_i) + M'_{l_1 - s_1, l_2 - s_2}(\tilde{p}_f, \tilde{q}''_i) \quad (64)$$
$$\times \frac{\hat{\tilde{q}}''_i + m_*}{\tilde{q}''^2_i - m^2_*} K''_{s_1 s_2}(\tilde{q}''_i, \tilde{p}_i) + K''_{s_1 s_2}(\tilde{p}_f, \tilde{q}''_f)$$
$$\left. \times \frac{\hat{\tilde{q}}''_f + m_*}{\tilde{q}''^2_f - m^2_*} M'_{l_1 - s_1, l_2 - s_2}(\tilde{q}''_f, \tilde{p}_i) \right].$$

Here, we introduced the following notations

$$K'_{s_1 s_2}(\tilde{p}_2, \tilde{p}_1) = \hat{\varepsilon}^* J_{s_1 s_2} + \frac{\omega_1 m^2}{8(k_1 \tilde{p}_1)(k_1 \tilde{p}_2)} B'_{s_1 s_2} \hat{k}_1, \quad (65)$$

$$K''_{s_1 s_2}(\tilde{p}_2, \tilde{p}_1)$$
$$= \frac{m}{4(k_1 \tilde{p}_1)} \hat{\varepsilon}^* \hat{k}_1 \hat{D}''_{s_1 s_2} + \frac{m}{4(k_1 \tilde{p}_2)} \hat{D}''_{s_1 s_2} \hat{k}_1 \hat{\varepsilon}^*, \quad (66)$$

$$M'_{l_1 = s_1, l_2 - s_2}(\tilde{p}_2, \tilde{p}_1) = \tilde{\gamma}_0 J_{l_1 = s_1, l_2 - s_2}$$
$$+ \frac{\omega_1 m^2}{8(k_1 \tilde{p}_1)(k_1 \tilde{p}_2)} B'_{l_1 = s_1, l_2 - s_2} \hat{k}_1, \quad (67)$$

$$M''_{l_1 - s_1, l_2 - s_2}(\tilde{p}_2, \tilde{p}_1) = \frac{m}{4(k_1 \tilde{p}_1)} \tilde{\gamma}_0 \hat{k}_1 \hat{D}''_{l_1 - s_1, l_2 - s_2}$$
$$+ \frac{m}{4(k_1 \tilde{p}_2)} \hat{D}''_{l_1 - s_1, l_2 - s_2} \hat{k}_1 \tilde{\gamma}_0. \quad (68)$$

The functions $B'_{s_1 s_2}$ and the four-vector $D''_{s_1 s_2}$ involved in expressions (65)–(68) are written as

$$B'_{s_1 s_2} = \eta_1^2 [J_{s_2 + 1, s_2 + 1} + J_{s_1 - 1, s_2 - 1} + 2J_{s_1 s_2}]$$
$$+ \eta_2^2 [J_{s_1 + 1, s_2 - 1} + J_{s_1 - 1, s_2 + 1} + 2J_{s_1 s_2}] \quad (69)$$
$$+ 2\eta_1 \eta_2 [J_{s_1 - 1, s_2} + J_{s_1 + 1, s_2} + J_{s_1, s_2 + 1} + J_{s_1, s_2 - 1}],$$

$$D''_{s_1 s_2} = e_x [\eta_1 (J_{s_1, s_2} + J_{s_1 - 1, s_2 - 1})$$
$$+ \eta_2 (J_{s_1, s_2 - 1} + J_{s_1 - 1, s_2})]. \quad (70)$$

We emphasize that the first two terms under the sum signs in (63) and (64) describe virtual processes of emission and absorption of an integer number of photons at combination frequencies ($\omega_1 \pm \omega_2$), whereas the last two terms are responsible for the virtual processes of emission and absorption of a half-integer number of photons at combination frequencies. Note also that the operator governing emission and absorption of a half-integer number of photons at combination frequencies, which is proportional to the intensities of external fields, $H^{\text{half-ind}}_{l_1 l_2} \sim \eta_{1,2}$, vanishes in the limiting case when $\eta_{1,2} \to 0$. In contrast to this operator, the expression for the operator $H^{\text{int}}_{l_1 l_2}$ in the limiting case of weak fields, $\eta_{1,2} \to 0$, is reduced to the formula that describes electron–nucleus SB in the absence of external fields. The amplitudes $S^{\text{int}}_{l_1 l_2}$ and $S^{\text{half-ind}}_{l_1 l_2}$ contribute to partial cross sections of electron–nucleus SB accompanied by emission (absorption) of an integer ($d\sigma^{\text{int}}_{l_1 l_2}$) and a half-integer ($d\sigma^{\text{half-ind}}_{l_1 l_2}$) number of photons at combination frequencies.

It is of interest to consider electron–nucleus SB in the plane perpendicular to the polarization vector in the presence of a single wave. As shown below, energy conservation in this case also allows us to distinguish between two situations: emission (absorption) of an even and an odd number of photons of the wave. Suppose, for example, that $F_2 = 0$. In this case, we have $\beta_2 = \alpha_+ = \alpha_- = 0$, and the expression for the amplitude defined by (62)–(70) can be written as

$$S_{fi} = \sum_{s_1 = -\infty}^{\infty} [S^{\text{even}}_{l_1} + S^{\text{odd}}_{l_1}], \quad (71)$$

where $S^{\text{even}}_{l_1}$ and $S^{\text{odd}}_{l_1}$ are the partial amplitudes corresponding to the emission (absorption) of an even and an odd number of photons of the wave, respectively. These amplitudes are determined by expression (8) where the replacements $q \longrightarrow q'$, $H_{ls} \longrightarrow H_{l_1}^{\text{even}}$ and $q \longrightarrow q''$, $H_{ls} \longrightarrow H_{l_1}^{\text{odd}}$ should be made. Then, the sum in $s_2$ vanishes in expressions (63) and (64), and the following replacements should be made:

$$K'_{s_1 s_2}(\tilde{p}_2, \tilde{p}_1) \longrightarrow K'_{s_1}(\tilde{p}_2, \tilde{p}_1) = \hat{\varepsilon}^* J_{s_1}(\beta_1) \\ + \frac{\omega_1 m^2}{8(k_1 \tilde{p}_1)(k_1 \tilde{p}_2)} B'_{s_1} \hat{k}_1, \quad (72)$$

$$K''_{s_1 s_2}(\tilde{p}_2, \tilde{p}_1) \longrightarrow K''_{s_1}(\tilde{p}_2, \tilde{p}_1) \\ = \frac{m}{4(k_1 \tilde{p}_1)} \hat{\varepsilon}^* \hat{k}_1 \hat{D}''_{s_1} + \frac{m}{4(k_1 \tilde{p}_2)} \hat{D}''_{s_1} \hat{k}_1 \hat{\varepsilon}^*, \quad (73)$$

$$M'_{l_1 - s_1, l_2 - s_2}(\tilde{p}_2, \tilde{p}_1) \longrightarrow M'_{l_1 - s_1}(\tilde{p}_2, \tilde{p}_1) \\ = \tilde{\gamma}_0 J_{l_1 - s_1}(\beta_1) + \frac{\omega_1 m^2}{8(k_1 \tilde{p}_1)(k_1 \tilde{p}_2)} B'_{l_1 - s_1} \hat{k}_1, \quad (74)$$

$$M''_{l_1 - s_1, l_2 - s_2}(\tilde{p}_2, \tilde{p}_1) \longrightarrow M''_{l_1 - s_1}(\tilde{p}_2, \tilde{p}_1) \\ = \frac{m}{4(k_1 \tilde{p}_1)} \tilde{\gamma}_0 \hat{k}_1 \hat{D}''_{l_1 - s_1} + \frac{m}{4(k_1 \tilde{p}_2)} \hat{D}''_{l_1 - s_1} \hat{k}_1 \tilde{\gamma}_0. \quad (75)$$

The functions $B'_{s_1}$ and the four-vector $D''_{s_1}$ in expressions (72)–(75) are given by

$$B'_{s_1} = \eta_1^2 [J_{s_1+1}(\beta_1) + J_{s_1-1}(\beta_1) + 2J_{s_1}(\beta_1)], \\ D''_{s_1} = e_x \eta_1 [J_{s_1}(\beta_1) + J_{s_1-1}(\beta_1)]. \quad (76)$$

The four-momenta defined by (56) and (57) can be represented as

$$\begin{cases} q' = \tilde{p}_f - \tilde{p}_i - k' + 2l_1 k_1 \\ \tilde{q}'_i = \tilde{p}_i - k' - 2s_1 k_1 \\ \tilde{q}'_f = \tilde{p}_f + k' + 2s_1 k_1, \end{cases} \\ \begin{cases} q'' = \tilde{p}_f - \tilde{p}_i - k' + (2l_1 - 1)k_1 \\ \tilde{q}''_i = \tilde{p}_i - k' - (2s_1 - 1)k_1 \\ \tilde{q}''_f = \tilde{p}_f + k' + (2s_1 - 1)k_1. \end{cases} \quad (77)$$

As can be seen from expressions (71)–(77), multiphoton processes accompanying electron–ion scattering and emission of a spontaneous photon in the plane perpendicular to the polarization vector (39) in the case of a single wave are determined by the quantum parameter $\beta_1$ (23). The processes that occur with emission (absorption) of an even and an odd number of photons of the wave are separated from each other in this case, as opposed to the range (40), where the Bunkin–Fedorov quantum parameter $\gamma_1$ (22) is the main multiphoton parameter.

Let us consider again electron–nucleus SB in the field of two waves in the interference range. Multiphoton processes in this range are described by the functions $J_{l_1 - s_1, l_2 - s_2}$ (41), and quantum parameters $\alpha_+$ and $\alpha_-$ (24) serve as the main multiphoton parameters. Therefore, electron–nucleus scattering and spontaneous photon emission in the interference range are mainly accompanied by the emission and absorption of $(l_1 < \alpha_+)$ photons at the combination frequency $(\omega_1 + \omega_2)$ and $(l_2 < \alpha_-)$ photons at the combination frequency $(\omega_1 - \omega_2)$. Hence, the fraction of energy emitted or absorbed by an electron in the initial or final state to or from both waves can be estimated in its order of magnitude as $l_1(\omega_1 + \omega_2)/E_{i,f} \lesssim \zeta_{i,f}$ or $l_2(\omega_1 - \omega_2)/E_{i,f} \lesssim \zeta_{i,f}$ (in the nonrelativistic limiting case, we have $2l_1(\omega_1 + \omega_2)/m v_{i,f}^2 \lesssim \zeta_{i,f}$ or $2l_2(\omega_1 - \omega_2)/m v_{i,f}^2 \lesssim \zeta_{i,f}$), where $\zeta_{i,f}$ is the classical interference parameter, which determines the integral characteristics of the process under study in the range defined by (39),

$$\zeta_{i,f} = \xi_1 \xi_2 \frac{p_{i,f}}{E_{i,f}}. \quad (78)$$

Here, the quantities $\xi_{1,2}$ are defined by expression (44) for the initial and final states. Let us specify the range of moderately strong fields within the interference range by imposing the following restriction on the classical interference parameter $\zeta_{i,f} \ll 1$. This restriction gives the following inequalities for the product of the intensities of the light waves:

$$\eta_1 \eta_2 \ll \begin{cases} v_{i,f}, & \text{if } v_{i,f} \ll 1 \\ 1, & \text{if } E_{i,f} \sim m \\ (E_{i,f}/m)^2, & \text{if } E_{i,f} \gg m. \end{cases} \quad (79)$$

Note that conditions (79) are less stringent than analogous conditions (45) in the noninterference range. By virtue of (79), we can neglect the energies of photons at combination frequencies as compared with the electron energy in expressions (56) and (57) for the squares of the four-vectors. Thus, we have $\tilde{q}_i^2 \neq m_*^2$ and $\tilde{q}_f^2 \neq m_*^2$. Consequently, the range of moderately strong fields (79) is nonresonant. If $\eta_1 \sim \eta_2$ in the range defined by (79), then we have $\tilde{p}_{i,f} = p_{i,f}$, and expressions (56) and (57) for the four-momenta are reduced to formula (46). In addition, we can neglect the quantity $H_{l_1 l_2}^{\text{half-ind}}$ (64). In expression (63) for $H_{l_1 l_2}^{\text{int}}$, we can neglect $B'$ and $D''$ (76) and perform summation over $s_1$ and $s_2$. Then, we find that

$$S_{l_1 l_2}^{\text{half-ind}} = 0, \quad S_{l_1 l_2}^{\text{int}} = J_{l_1 l_2}(\beta_1, \beta_2; \alpha_+, \alpha_-) S_*, \quad (80)$$

where $S_*$ is the electron–nucleus SB amplitude in the absence of an external field [22] and the functions $J_{l_1 l_2}$ are defined by expression (41) where the arguments (23) and (24) should be taken with $\tilde{p}_1 = p_i$ and $\tilde{p}_2 = p_f$, i.e.,

$$\beta_{1,2} = \beta_{1,2}(p_f, p_i), \quad \alpha_\pm = \alpha_\pm(p_f, p_i). \quad (81)$$

Analyzing expression (80) for the amplitude, we can easily find the partial differential cross section of electron–nucleus SB in the field of two waves within the interference range

$$d\sigma_{l_1 l_2}^{\text{int}} = |J_{l_1 l_2}|^2 d\sigma_*. \quad (82)$$

Here, $d\sigma_*$ is the differential cross section of electron–nucleus SB in the absence of an external field [22]. We emphasize that the integer indices $l_1$ and $l_2$ in the partial cross section (82) correspond to the emission (absorption) of an integer number of photons at combination frequencies $(\omega_1 + \omega_2)$ and $(\omega_1 - \omega_2)$, as opposed to a similar expression (50) for the noninterference range, where integer indices $l$ and $s$ correspond to the emission (absorption) of mean numbers of photons of the first and second waves. Note that, within the range of wave intensities that satisfy conditions opposite of (51), i.e.,

$$\eta_{1,2}^2 \gtrsim \begin{cases} \omega_{1,2}/m v_{i,f}, & \text{if } v_{i,f} \ll 1 \\ \omega_{1,2}/m, & \text{if } E_{i,f} \sim m \\ \omega_{1,2} E_{i,f}/m^2, & \text{if } E_{i,f} \gg m, \end{cases}$$

$$\eta_1 \eta_2 \gtrsim \begin{cases} \omega_{1,2}/m v_{i,f}, & \text{if } v_{i,f} \ll 1 \\ \omega_{1,2}/m, & \text{if } E_{i,f} \sim m \\ \omega_{1,2} E_{i,f}/m^2, & \text{if } E_{i,f} \gg m, \end{cases} \quad (83)$$

we have $\beta_{1,2} \gtrsim 1$ and $\alpha_\pm \gtrsim 1$, and, consequently, $\gamma_{1,2} \gg \alpha_\pm \gtrsim 1$ and $\gamma_{1,2} \gg \beta_{1,2} \gtrsim 1$ [see (43), (49), and (81)]. Taking these relations into account, we can easily demonstrate [31] that the partial cross section (50) in the noninterference range for wave intensities, meeting conditions (45) and (83), is much less than the corresponding cross section (82) in the interference range

$$\frac{d\sigma_{ls}}{d\sigma_{l_1 l_2}^{\text{int}}} = \frac{|I_{ls}|^2}{|J_{l_1 l_2}|^2}$$

$$\sim \begin{cases} (\gamma_1 \gamma_2)^{-1} \ll 1, & \text{for } l_1 \sim l \ll \gamma_1, \ l_2 \sim s \ll \gamma_2 \\ (\gamma_1 \gamma_2)^{-2/3} \ll 1, & \text{for } l_1 \sim l \sim \gamma_1, \ l_2 \sim s \sim \gamma_2. \end{cases} \quad (84)$$

Summing the partial cross sections (82) for all the processes of emission and absorption of photons at combination frequencies, we find that, similar to the noninterference range, all the essentially quantum terms cancel each other out [see (53)]

$$d\sigma^{\text{int}} = \sum_{l_1 = -\infty}^{\infty} \sum_{l_2 = -\infty}^{\infty} d\sigma_{l_1 l_2}^{\text{int}} = d\sigma_*. \quad (85)$$

Note also that, in the case of a single wave, expression (82) is reduced to a formula for the partial cross section of electron–nucleus SB with emission (absorption) of an even number of wave photons [the partial cross section with emission or absorption of an odd number of wave photons is equal to zero in this case, see (71)–(77)]

$$d\sigma_{l_1}^{\text{even}} = J_{l_1}^2(\beta_1) d\sigma_*. \quad (82)$$

## 6. CONCLUSION

The performed investigation of SB that accompanies the scattering of an electron by a nucleus in the field of two linearly polarized light waves shows that

(1) The process of electron–nucleus SB in the field of two waves is highly sensitive to the kinematics of electron scattering and spontaneous photon emission, which makes it possible to distinguish between two kinematic ranges: the noninterference range (40), where the Bunkin–Fedorov quantum parameters (22) serve as the main multiphoton parameters, and the interference range (39), where the interference quantum parameters $\alpha_\pm$ (24) play the role of multiphoton parameters;

(2) For moderately strong fields (45), the partial cross section within the noninterference range can be factorized as a product of the probability to emit (absorb) a certain number of photons of the first and second waves and the cross section of electron–nucleus SB in the absence of an external field (50);

(3) The interference range is characterized by electron scattering and spontaneous photon emission in the plane perpendicular to the polarization vectors of light waves. In this case, an electron emits and absorbs an integer or a half-integer number of photons at combination frequencies $(\omega_1 \pm \omega_2)$ in electron–nucleus bremsstrahlung and emission of a spontaneous photon. The processes of emission and absorption of integer and half-integer numbers of photons at combination frequencies are separated by the energy-conservation law;

(4) For moderately strong fields (79), the partial cross section within the interference range can be factorized as a product of the probability to emit (absorb) an integer number of photons at combination frequencies and the cross section of electron–nucleus SB in the absence of an external field (82);

(5) For wave intensities that satisfy conditions (45) and (83), the partial cross section (82) in the interference range is much larger than the corresponding cross section (50) in the noninterference range.